\documentclass[aps,showkeys,showpacs,nosuperscriptaddress,twocolumn]{revtex4}

\usepackage{psfrag}
\usepackage{graphicx}
\usepackage{amsmath}
\usepackage{hyperref}
\usepackage{color}

\begin{document}

\title{Strategies in crowd and crowd structure}

\author{P.~Gawro\'nski}
\email{gawron@newton.ftj.agh.edu.pl}
\affiliation{
AGH University of Science and Technology, Faculty of Physics and Applied Computer Science, al. Mickiewicza 30, PL-30059 Krakow, Poland
}

\author{K.~Malarz}
\email{malarz@agh.edu.pl}
\affiliation{
AGH University of Science and Technology, Faculty of Physics and Applied Computer Science, al. Mickiewicza 30, PL-30059 Krakow, Poland
}

\author{M.~J.~Krawczyk}
\email{gos@fatcat.agh.edu.pl}
\affiliation{
AGH University of Science and Technology, Faculty of Physics and Applied Computer Science, al. Mickiewicza 30, PL-30059 Krakow, Poland
}

\author{J.~W\c{a}s}
\email{jarek@agh.edu.pl}
\affiliation{
AGH University of Science and Technology, Faculty of Electrical Engineering, Automatics, Computer Science and Engineering in Biomedicine, al. Mickiewicza 30, PL-30059 Krakow, Poland
}

\author{J.~Malinowski}
\email{Janusz.Malinowski@fis.agh.edu.pl}
\affiliation{
AGH University of Science and Technology, Faculty of Physics and Applied Computer Science, al. Mickiewicza 30, PL-30059 Krakow, Poland
}

\author{A.~Kupczak}
\email{kupczak.a@gmail.com}
\affiliation{
AGH University of Science and Technology, Faculty of Physics and Applied Computer Science, al. Mickiewicza 30, PL-30059 Krakow, Poland
}

\author{W.~Sikora}
\email{wsikora@agh.edu.pl}
\affiliation{
AGH University of Science and Technology, Faculty of Physics and Applied Computer Science, al. Mickiewicza 30, PL-30059 Krakow, Poland
}

\author{J.~W.~Kantelhardt}
\email{jan.kantelhardt@physik.uni-halle.de}
\affiliation{
Martin-Luther-Universit{\"a}t Halle-Wittenberg, Institut f{\"u}r Physik, von-Seckendorff-Platz 1, D-06099 Halle/Saale, Germany
}

\author{K.~Ku{\l}akowski}
\email{kulakowski@fis.agh.edu.pl}
\affiliation{
AGH University of Science and Technology, Faculty of Physics and Applied Computer Science, al. Mickiewicza 30, PL-30059 Krakow, Poland
}

\date{\today}

\begin{abstract}
In an emergency situation, imitation of strategies of neighbours can lead to an order-disorder phase transition, where spatial clusters
of pedestrians adopt the same strategy. We assume that there are two strategies, cooperating and competitive, which correspond to a smaller or larger desired velocity. The results of our simulations within the Social Force Model indicate that the ordered phase can be detected as an increase of spatial order of positions of the pedestrians in the crowd.
\end{abstract}



\maketitle

%% ###########################################################################
\section{Introduction}
\label{S1}
%% ###########################################################################

Simulations of crowd dynamics is among the most actively developed applications of physical methods in social sciences. Standard difficulties met in such applications are in this case either absent, or seriously weakened. First, crowd dynamics is relevant for control of large gatherings and prevention of stampede catastrophes  \cite{stam,stamm}. Second, experimental data are accessible \cite{data,data2}. Third, a model (Social Force Model, SFM) has been developed \cite{sfm} which can be used by physicists without any input of social sciences. Fourth, realistic values of the model parameters have been obtained and verified \cite{params}. These causes made SFM a popular tool, used and developed by various scientific groups of physicists and computer scientists. To develop its connections with collective sociological effects is one of our motivations here.

In the original version of SFM \cite{sfm,params}, most of the parameters are related to physical aspects of the interactions between pedestrians and of the interactions of pedestrians with obstacles. There are two exceptions: the strength of interpersonal, psychologically motivated repulsion $A$ and the desired velocity $v_d$. A variation of $A$ has been used in \cite{lnai,cpc} to include individual decisions on calling for help and coming to aid. Here we are interested in a variation of the desired velocity $v_d$. In SFM, equations of motion for each pedestrian are solved numerically, and one of the forces is proportional to the difference $\bf v_d-\bf v(t)$, where $\bf v(t)$ is the actual velocity of the given pedestrian. In the conditions of evacuation, $v_d$=1.8 m/s seems to be appropriate as an accepted velocity of brisk walking \cite{brisk}. On the contrary, $v_d$=3.0 m/s was found to produce jamming at narrow exits \cite{app}. Here we use these two values to differentiate between two strategies of evacuation:
 cooperating (1.8 m/s) and competitive (3.0 m/s).

The collective character of the process is due to the imitation effect. Namely, we use probabilities $w(n)$ that the strategy (cooperative or competitive, $X=C$ or $S$) of a given pedestrian will be changed if $n$ out of 6 direct neighbours of a given pedestrian adopt this strategy. The number six is the number of nearest neighbours in a triangular two-dimensional lattice, what is a model of people in crowd; suggested by our former observations \cite{app}. In our former work, we used a square lattice  \cite{rudi}; yet we do not expect qualitative differences between the results for these two structures.

Our aim here is as follows. We expect that the imitation leads to spatial correlations of strategies of pedestrians, and that these correlations can lead to a respective phase transition \cite{rudi}. We ask if this phase transition is visible in the crowd structure. This interest is motivated by the idea of detecting collective effects online by analysis of crowd motion, registered by local cameras.

The strategy of our modelling is as follows. First we set the probabilities $w(n)$ as dependent on a single parameter $x$ and we solve numerically the Ising-like problem of selecting strategies $C$ or $S$ by pedestrians. As the outcome we get the set ${w_c(n)}$ where there is a transition between the ordered and the disordered phase. Next we verify the obtained transition point by means of the fluctuation function  \cite{jan}. These steps are described in Section \ref{S2}. Next we perform the simulations of crowd dynamics within SFM, where pedestrians select $v_d$= 1.8 m/s (strategy $C$) or 3.0 m/s (strategy $S$). During the simulation, we evaluate how a local crowd structure depends on the parameter $x$ and, in particular, we check if the above mentioned phase transition is visible. This step is described in Section \ref{S3}. These considerations are illustrated by the results of our field observations of crowd, described in Section \ref{S4}. Last section is devoted to discussion.

%% ###########################################################################
\section{The phase transition}
\label{S2}
%% ###########################################################################

At this preliminary stage of modelling pedestrians are placed at nodes of a two-dimensional, triangular lattice and their motion is not taken into account. The lattice contains $N=10^6$ sites and helical boundary conditions are assumed. In the initial state, strategies $C$ and $S$ are assigned to pedestrians with given probabilities $r$, $1-r$; spatial correlations are not taken into account. During the simulation, pedestrians change their strategies with some probabilities $w(n)$, that depend on the number $n$ of neighbouring subjects currently using the same strategy (see also \cite{cpc}). Here the functions $w(n)$ depend on a single control parameter $x$ as follows: $w(0)=1$, $w(1)=3x$, $w(2)=2x$, $w(3)=x$, $w(4)=x/2$, $w(5)=x/4$, $w(6)=x/6$.  If $w(n)>1$ for any $n$, we set $w(n)=1$. For simplicity, these probabilities are the same for both strategies $C$ and $S$.

\begin{figure}[!hptb]
\begin{center}
\psfrag{m}{$\langle m\rangle$}
\psfrag{x}{$x$}
\includegraphics[width=\columnwidth]{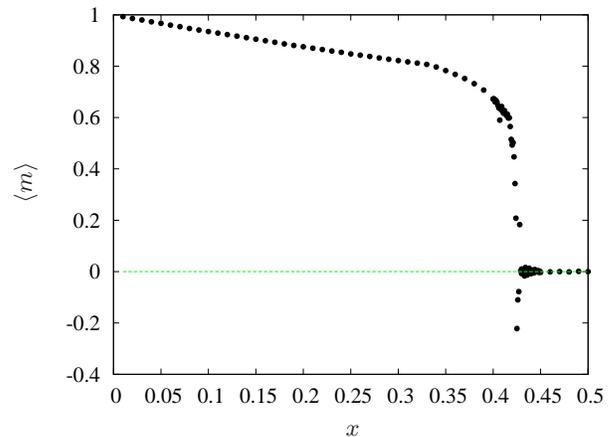}
\caption{Average order parameter $\langle m\rangle$ dependence on control parameter $x$. The last $M/2=5\cdot 10^7$ steps of $m(t)$ temporal evolution is used for average $\langle m\rangle$ evaluation.}
\label{fig1}
\end{center}
\end{figure}

\begin{figure}[!hptb]
\begin{center}
\psfrag{x=}{$x=$}
\psfrag{m}{$m$}
\psfrag{t/1E8}{$t/10^8$}
\includegraphics[width=\columnwidth]{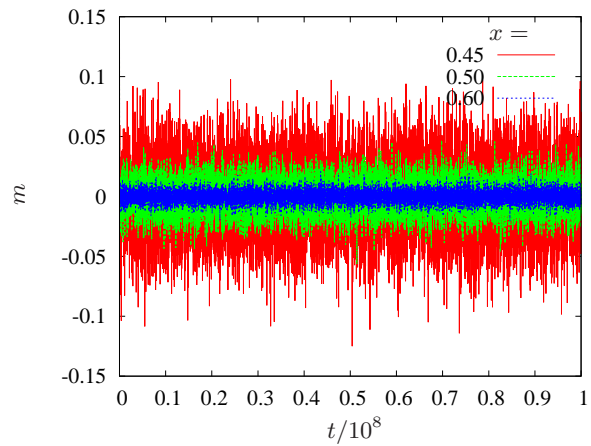}
\caption{(Colour online) The temporal dependence of order parameter $m(t)$ for several values of the control parameter $x=0.45$, 0.5 and 0.6 ($x>x_c$). The range of fluctuations increase when we approach the transition point $x_c$ from above.}
\label{fig2}
\end{center}
\end{figure}

The numerical results indicate that indeed an order-disorder phase transition appears near $x_c=0.429$, as shown in Fig. \ref{fig1}. For $x<x_c$, the probabilities of state changes are small and clusters of pedestrians with the same strategy persist. This result is confirmed by the analysis of temporal dependence of the order parameter, defined as $m=(N(C)-N(S))/N$; there, $N(X)$ is the number of agents who adopt strategy $X$, and $N=N(C)+N(S)=10^6$. The temporal dependence of $m$ for several control parameters $x=0.45$, 0.5 and 0.6 ($x>x_c$) is presented on Fig. \ref{fig2}. The simulation takes $M=10^8$ steps and single step is completed when all $N=10^6$ nodes are investigated in typewriter order. The last half of these steps ($M/2=5\cdot 10^7$) is used for average order parameter $\langle m\rangle$ evaluation (Fig. \ref{fig1}). 

The continuous character of the phase transition is confirmed by studying the persistence of order parameter time series $m(t)$ using multifractal detrended fluctation analysis with linear detrending (MF-DFA1) \cite{jan}. In this approach, the $q$-th order fluctuation function is defined as

\[ F_q(s)=\left\{\dfrac{1}{2N_s}\sum\limits_{\nu=1}^{2N_s}\left[F^2(\nu,s)\right]^{q/2}\right\}^{1/q}, \]
where $F^2(\nu,s)$ is the detrended variance calculated for each segment $\nu$ of the length $s$, which is obtained by the division of the analysed time series into non-overlapping segments of equal length; $q$ can be any real value, here we use $q\in\{ -2, 0.5, 2\}.$

\begin{figure}[!hptb]
\begin{center}
\includegraphics[width=\columnwidth]{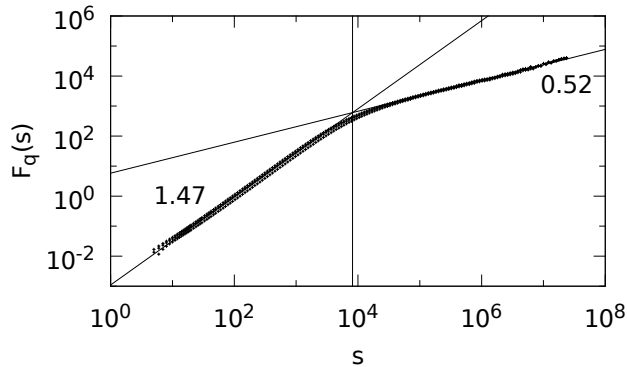}
\caption{MF-DFA fluctuation functions $F_q(s)$ versus the scale $s$ for $q\in\{-2, 0.5, 2\}$; as it is seen, the curves for different $q$'s almost overlap.}
\label{fig3a}
\end{center}
\end{figure}

\begin{figure}[!hptb]
\begin{center}
\includegraphics[width=\columnwidth]{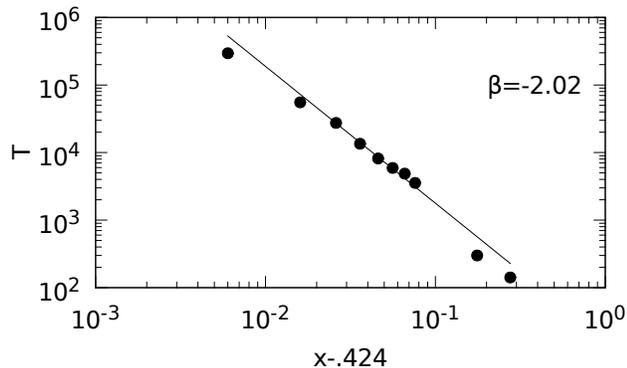}
\caption{The temporal range $T$ of correlations as dependent on $x$.}
\label{fig3b}
\end{center}
\end{figure}

Figure \ref{fig3a} shows typical MF-DFA fluctuation functions for $x\approx 0.42$.  One can see a crossover from random walk type scaling behaviour ($F_q(s) \sim s^{h(q)}$ with $h(q) \approx 1.5$) on small time scales $s$ to random white noise type scaling behaviour ($h(q) \approx 0.5$) on large time scales. The temporal range $T$ of the correlations is given by the position of the crossover, being close to $10^4$ in this case.  Figure \ref{fig3b} shows the dependence of $T$ on $x$, which follows a power law, $T(x) \sim (x-x_c)^{\beta}$ with $\beta \approx -2$.  We see that $T$ continously diverges in the vicinity of the transition point.

%% ###########################################################################
\section{The crowd simulations}
\label{S3}
%% ###########################################################################

The simulation is performed in accordance to the original formulation of the Social Force Model and the model parameters \cite{sfm,data}.  Pedestrians are moving, all in the same direction, along a corridor of five meters width, with periodic boundary conditions. After each one second, they check the strategies adapted at the moment by their six nearest neighbours, and they modify their strategies according to the probabilities $w(n)$, the same as given in Section \ref{S2}.

To inspect the crowd structure, for each pedestrian $i$ we identify his six nearest neighbours and we calculate the difference $D_i$ between the distances from $i$ to his nearest neighbour and from $i$ to his sixth nearest neighbour. In the triangular lattice, the difference is zero. The smaller $D_i$, the crowd is more ordered.

\begin{figure}[!hptb]
\begin{center}
\psfrag{< D >}{$\langle D\rangle$}
\psfrag{X}{$x$}
\includegraphics[width=\columnwidth]{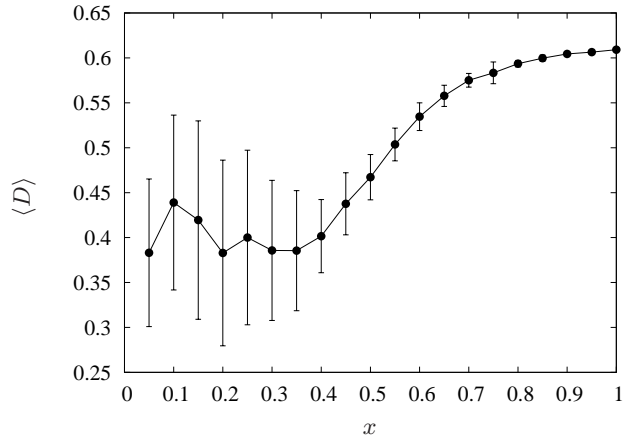}
\caption{The mean difference of distances $\langle D\rangle$ as dependent on $x$, taken from the SFM simulations.}
\label{fig4}
\end{center}
\end{figure}

\begin{figure}[!hptb]
\begin{center}
\psfrag{< r >}{$\langle r\rangle$}
\psfrag{X}{$x$}
\includegraphics[width=\columnwidth]{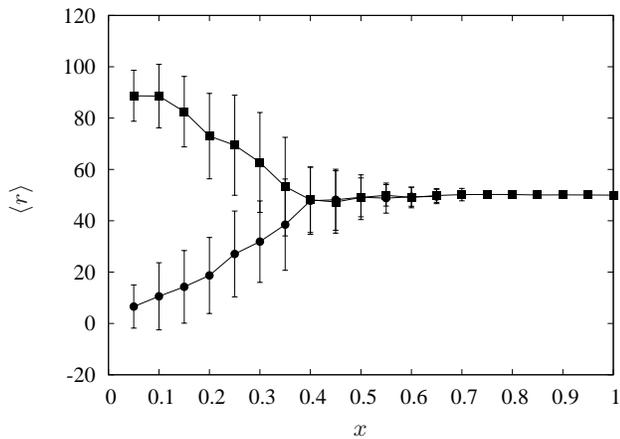}
\caption{The mean value $\langle r\rangle$ of the percentage of the pedestrians who adopt the competitive strategy, taken from the SFM simulations. Below $x_c$, the results depend on the initial value of $r$, which was taken as 0.7 (upper curve) and 0.3 (lower curve). }
\label{fig5}
\end{center}
\end{figure}

In Fig. \ref{fig4}, we show the dependence of the mean difference of distances $\langle D\rangle$ on the parameter $x$, measured at the tentatively stationary state. The mean is calculated over all pedestrians $i=1,\cdots,N$ and over a time period of about ten seconds, after a transient time of the same length. As we see, $\langle D\rangle$ is larger but more stable above the transition point, i.e. for $x>x_c$. Below this value of $x$, a clear reduction of $\langle D\rangle$ can be observed, although large fluctuations hamper the accuracy. On the other hand, the information on the initial state, encoded by the probability $r$, is preserved below the transition; this is shown in Fig. \ref{fig5}.

%% ###########################################################################
\section{Field observation}
\label{S4}
%% ###########################################################################

The Corpus Christi procession in Krakow, Grodzka Street, was registered by a hand camera on June 7, 2012 from a window at 1-st floor, about five meters above the street. The movie lasts ten minutes, but after about first five minutes and 30 seconds the crowd density started to decrease. Also, during two first minutes priests and military orchestra are moving in a prescribed order. Then the material is limited to about 3'30'', when the crowd density is about one person per m$^2$.

\begin{figure}[!hptb]
\begin{center}
\includegraphics[width=\columnwidth]{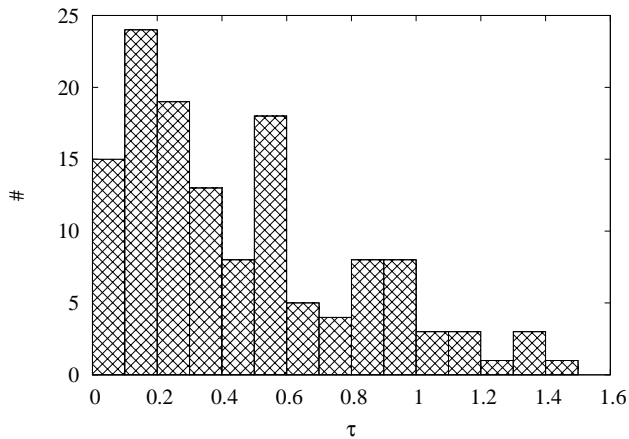}
\caption{The experimental histogram of time gaps $\tau$ between pedestrians, as taken from the movie.}
\label{fig6}
\end{center}
\end{figure}

The aim was to check if some structure of the crowd can be found. For this purpose, a lane along the street is selected of about three meters width. For pedestrians who appear at this lane, the times when their heads disappear from the screen are registered manually. The differences between these times are used to make a histogram of time gaps $\tau$, shown in Fig. \ref{fig6}. We note that the velocity of the crowd is more or less constant.

We expected and hoped to see two peaks in the histogram of the time gaps $\tau$, and indeed the obtained plot does not shatter these expectations. First peak, at about 0.2 s, is seen in the plots for all movies we did, and it can be safely interpreted as related to pedestrians who walk together. The second peak about 0.5 s is specific for this crowd density. Once the density decreases, the times between subsequent groups vary much more and the second peak is not visible.

%% ###########################################################################
\section{Discussion}
\label{S5}
%% ###########################################################################

The results of numerical simulations within SFM indicate, that if the imitation process leads to a phase transition, and if the two states are related with different values of the desired velocity, then the ordered phase manifests itself by an increase of short-range order. On the contrary, in the disordered phase the spatial correlations decrease, what means that clusters of people moving with the same velocity disappear.

It would be desirable to confirm this conclusion by our field observation. However, the data recorded are related only to an ordered phase, where pedestrians move with approximately the same velocity, and the statistics is rather poor. Then, the obtained spectrum of times between neighbouring pedestrians can serve here as merely an illustration that some amount of spatial short range order does exist.

Our results suggest, that it is possible to infer about the state of crowd from its spatial structure registered in one instant of time. More local ordering indicates the presence of collective effects, which can be an effect of imitation. This tool of crowd control can be complementary to correlations of velocity of pedestrians, detected from movies \cite{stam}.

%% ###########################################################################
\begin{acknowledgments}
The work was partially supported by the Polish Ministry of Science and Higher Education and its grants for Scientific Research, by the PL-Grid Infrastructure, and by the by the EU project SOCIONICAL (FP7, grant 231288).
\end{acknowledgments}
%% ###########################################################################

 \end{document}